# DESIGNING A CYBER-SECURITY CULTURE ASSESSMENT SURVEY TARGETING CRITICAL INFRASTRUCTURES DURING COVID-19 CRISIS

Anna Georgiadou, Spiros Mouzakitis, and Dimitris Askounis

Decision Support Systems Laboratory, National Technical University of Athens, Iroon Polytechniou 9, 15780 Zografou, Greece

## ABSTRACT

*The paper at hand presents the design of a survey aiming at the cyber-security culture assessment of critical infrastructures during the COVID-19 crisis, when living reality was heavily disturbed and working conditions fundamentally affected. The survey is rooted in a security culture framework layered into two levels, organizational and individual, further analyzed into 10 different security dimensions consisted of 52 domains. An in-depth questionnaire building analysis is presented focusing on the aims, goals, and expected results. It concludes with the survey implementation approach while underlining the framework's first application and its revealing insights during a global crisis.*

## 1. INTRODUCTION

Coronavirus disease 2019, widely known as COVID-19, is an infectious disease caused by severe acute respiratory syndrome coronavirus 2 (SARS-CoV-2) [1]. The disease was first detected in late 2019 in the city of Wuhan, the capital of China's Hubei province[2]. In March 2020, the World Health Organization (WHO) declared the COVID-19 outbreak a pandemic [3]. Today, with more than 11 million confirmed cases in 188 countries and at least half a million casualties, the virus is continuing its spread across the world. While epidemiologists argue that the crisis is not even close to being over, it soon becomes apparent that "the COVID-19 pandemic is far more than a health crisis: it is affecting societies and economies at their core" [4].

Terms such as "Great Shutdown" and "Great Lockdown" [5, 6, 7] have been introduced to attribute the major global recession which arose as an economic consequence of the ongoing COVID-19 pandemic. The first noticeable indication of the accruing recession was the 2020 stock market crash on the 20th of February. International Monetary Fund (IMF) in the April World Economic Outlook projected global growth in 2020 to fall to -3 percent. This is a downgrade of 6.3 percentage points from January 2020, making the "Great Lockdown" the worst recession since the Great Depression, and far worse than the Global Financial Crisis [7]. According to the International Labour Organization (ILO) Monitor, published on 7th April 2020, full or partial lockdown measures are affecting almost 2.7 billion workers, representing around 81% of the world's workforce [8].

Organizations from various business domains and operation areas globally try to survive this unprecedented financial crisis by investing their hopes, efforts, and working reality on information technology and digitalization. The workforce is being encouraged and facilitated on teleworking while most products and services become available over the web while, in many cases, transforming and adjusting to current rather demanding reality. However, the aforementioned organisations face another, not that apparent, COVID-19 side-effect: the cyber-crime increase.

The increase in the population percentage connected to the World Wide Web, the expansion of time spent online, combined with the sense of confinement and the anxiety and fear generated from the lockdown, seem to catalyzeaction of cyber-criminals. Coronavirus has rapidly reshaped the dark web activities, as buyers and sellers seize the opportunity to capitalizeon global fears, as well as dramatic shifts in supply and demand. Phishing emails, social engineering attacks, malware, ransom ware and spyware, medical related scums, investment opportunities frauds, are only a few examples of the cyber-crime incidents reported during the crisis period [9, 10].

INTERPOL's Cybercrime Threat Response team has detected and reported a significant increase in the number of attempted ransom ware attacks against key organizations and infrastructure engaged in the virus response. Cybercriminals are using ransom ware to hold hospitals and medical care services digitally hostage; preventing them from accessing vital files and systems until a ransom is paid[11].

Cyber-security agencies, organizations, and experts worldwide have issued recommendations and proposed safeguard measures to assist individuals and corporations defend against cyber-crime. While the virus is dominating in every aspect of our daily lives and human interaction is being substituted by digital transactions, cybersecurity gains the role it was deprived from during the last years. The question that remains unanswered, given the circumstances, is: What are the COVID-19 pandemic cyber-security culture side-effects on both individual and organizational level?

The manuscript at hand presents the design and rollout plan of a survey aiming to assess the cyber-security culture during the COVID-19 pandemic in the critical infrastructure domain. Section 2 presents background information regarding the importance of public cyber-security surveys conducted over the years, emphasizingon the variety and originality of their findings. Building upon their approach, a detailed methodology is reported in Sections 3 & 4, in an effort to develop a brief, targeted and comprehensible survey for the assessment of the cybersecurity readiness of organizations during the crisis with emphasis on employees' feelings, thoughts, perspective, individuality. In Section 5, we sketch the survey next steps towards its conduction and fruitful completion. Finally, Section 6 concludes by underlying the importance of our survey reasoning while focusing on the challenging scientific opportunities that arise from it.

## 2. BACKGROUND

Over the last decades, cybersecurity surveys have been a powerful asset to academics and information security professionals seeking to explore the ever-changing technological reality. Their goal was to uncover current trends in cyber threats, organizations' investment priorities, cloud security solutions, threat management, application security, security training and certification, and more.

Initially, they were narrowed down and addressed to certain participants depending on the nature and specific purpose of each survey. A lighthouse representative of this kind was the Computer Crime & Security Survey conducted by the Computer Security Institute (CSI) with the participation of the San Francisco Federal Bureau of Investigation's (FBI) Computer Intrusion

Squad. This annual survey, during its 15 years of life (starting from 1995 and reaching up to 2010), was probably one of the longest-running continuous surveys in the information security field[12]. This far-reaching study provided unbiased information and analysis about targeted attacks, unauthorized access, incident response, organizational economic decisions regarding computer security and risk management approaches based on the answers provided by computer security practitioners in U.S. corporations, government agencies, financial institutions, medical institutions and universities.

Following their lead, many public and private sector organizations are seeking revealing findings that will help them calibrate their operations and improve their overall presence in the business world via cybersecurity surveys. Healthcare Information and Management Systems Society (HIMSS) focusing on the health sector[13]; ARC Advisory Group targeting Industrial Control Systems (ICS) in critical infrastructures such as energy and water supply, as well as in process industries, including oil, gas and chemicals [14]; SANS exploring the challenges involved with the design, operation and risk management of ICS, its cyber assets and communication protocols, and supporting operations[15]; Deloitte in conjunction with Wakefield Research interviewing C-level executives who oversee cybersecurity at companies [16]; these being only some of the countless examples available nowadays.

Current trend in the cybersecurity surveys seems to be broadening their horizon by making them available and accessible through the internet [17, 18]. Since their goal is to reach out and attract more participants, thus enriching the collected data and, consequently, enforcing their results, tend to be shorter, more comprehensive to the majority of average users and apparently web-based.

Recognizing the unique value of this undisputable fruitful security evaluation methodology and rushing from the special working and living circumstances due to the COVID-19 pandemic, we identified the research opportuning to evaluate how this crisis has affected the cybersecurity culture of both individuals and organizations across the suffering globe. Security threats, frauds, breaches & perils have been brought to the light, recommendations have been given and precautions have been made [19, 20, 21]. What about the cybersecurity culture and its potential scars from this virus? Addressing this concern was our aim when designing, conducting and analyzing the survey presented in this paper.

## 3. SECURITY CULTURE FRAMEWORK

During the last years, our research efforts have been focusing on cyber-security in terms of tools, standards, frameworks and marketplace solutions especially targeting the human element. We have benchmarked the dominant reveals on the field, classified their possibilities and analyzed their core security factors. Having identified their gaps and overlaps, common grounds and differentiation and thoroughly studied several academic principles regarding information security, including technical analyses, algorithmic frameworks, mathematical models, statistical computations, behavioral, organizational and criminological theories, we have created a foundation combining the elements that constitute the critical cyber-security culture elements [22]. The suggested cybersecurity culture framework is based on a domain agnostic security model combining the key factors affecting and formulating the cybersecurity culture of an organization. It is layered into two levels, organizational and individual, analyzed into 10 different security dimensions consisted of 52 domains assessed by more than 500 controls. This hierarchical approach is being presented in Figure 1. Table 2 and Table 4list dimensions, domains and indicative controls in an attempt to unfold to the reader the generalized philosophy of our framework.

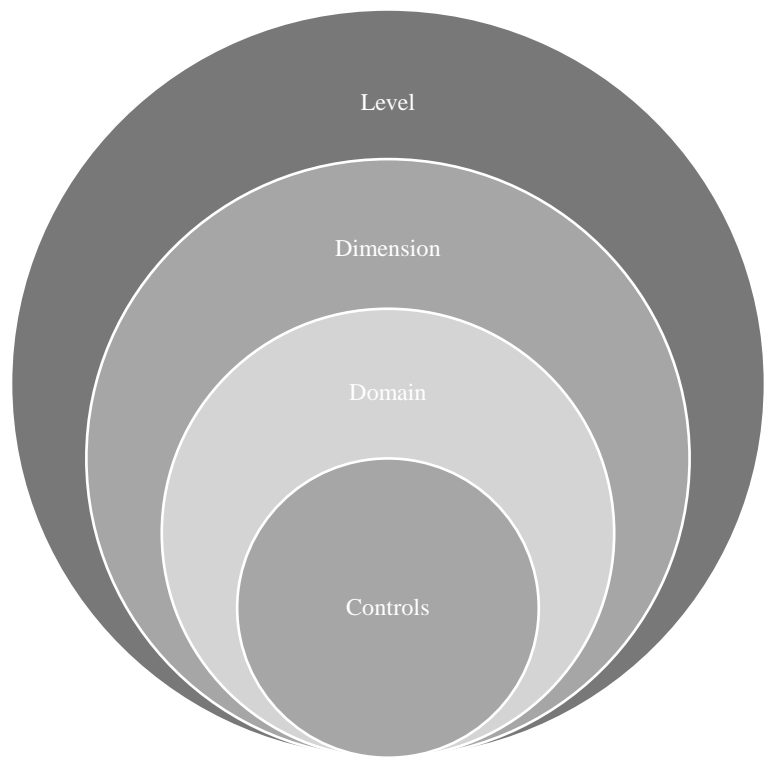

Figure 1. Cyber-Security Culture Model: Main Concepts

Table 1.Organisational Levelpresenting indicative controls

| Dimension | Domain | Indicative Controls |
|---|---|---|
| Assets | Application Software Security | <ul><li>Do you only use up-to-date and trusted third-party components for the software developed by the organization?</li><li>Do you apply static and dynamic analysis tools to verify that secure coding practices are being adhered to for internally developed software?</li></ul> |
| | Data Security and Privacy | <ul><li>Do you maintain an inventory of all sensitive information stored, processed, or transmitted by the organization's technology systems, including those located on-site or at a remote service provider?</li><li>Have you ensured that sensitive data or systems are not regularly accessed by the organization from the network?</li></ul> |
| | Hardware Assets Management | <ul><li>Do you employ integrity checking mechanisms to verify hardware integrity?</li><li>Do you maintain an accurate and up-to-date inventory of all assets with the potential to store or process information?</li></ul> |
| | Hardware Configuration Management | <ul><li>Have you established and do you maintain secure configuration management processes (e.g. when servicing field devices or updating their firmware)?</li><li>Do you store the master images and templates on securely configured servers, validated with integrity monitoring tools, to ensure that only authorized changes to the images are possible?</li></ul> |
| | Information Resources Management | <ul><li>Do you properly label all relevant assets, depending on their classification?</li><li>Are the classification scheme and labeling procedures properly communicated to all relevant parties?</li></ul> |

| | Network Configuration Management | ▪ Do you maintain documented security configuration standards for all authorized network devices?<br>▪ Have you compared all network device configurations against approved security configurations defined for each network device in use, and do you alert when any deviations are discovered? |
|---|---|---|
| | Network Infrastructure Management | ▪ Have you associated active ports, services, and protocols to the hardware assets in the asset inventory?<br>▪ Do you perform automated port scans on a regular basis against all systems and alert if unauthorized ports are detected on a system? |
| | Software Assets Management | ▪ Have you utilized software inventory tools throughout the organization to automate the documentation of all software on business systems?<br>▪ Is the software inventory system tied into the hardware asset inventory so that all devices and associated software are tracked from a single location? |
| | Personnel Security | ▪ Does your staff wear ID badges?<br>▪ Are authorized access levels and type (employee, contractor, visitor) identified on the Badge? |
| | Physical Safety and Security | ▪ Is access to your computing area controlled (single point, reception or security desk, sign-in/sign-out log, temporary/visitor badges)?<br>▪ Do you have an emergency evacuation plan and is it current? |
| **Continuity** | Backup Mechanisms | ▪ Do you store backups in a remote location?<br>▪ Do you encrypt backups containing confidential information? |
| | Business Continuity & Disaster Recovery | ▪ Do you have an emergency/incident management communications plan?<br>▪ Do you have a current business continuity plan? |
| | Capacity Management | ▪ Do you have enough capacity to ensure that data availability is maintained?<br>▪ Do you either deny or restrict bandwidth for resource-hungry services if these are not business critical? |
| | Change Management | ▪ Are the maintenance and copying of program source libraries subject to strict change control?<br>▪ Have you established a formal approval procedure for proposed changes? |
| | Continuous Vulnerability Management | ▪ Do you perform authenticated vulnerability scanning with agents running locally on each system or with remote scanners that are configured with elevated rights on the system being tested?<br>▪ Have you utilized a risk-rating process to prioritize the remediation of discovered vulnerabilities? |
| **Access and Trust** | Access Management | ▪ Have you enabled firewall filtering between VLANs to ensure that only authorized systems are able to communicate with other systems necessary to fulfill their specific responsibilities?<br>▪ Have you implemented physical or logical access controls for the isolation of sensitive applications, application data or systems? |

|  | Account Management | ▪ Do you automatically disable dormant accounts after a set period of inactivity?<br>▪ Do you maintain an inventory of each of the organization's authentication systems, including those located on-site or at a remote service provider? |
|---|---|---|
|  | Communication | ▪ Do you have documentation of the mapping of organizational communication flows?<br>▪ Do users acknowledge receipt of secret authentication information? |
|  | External Environment Connections | ▪ Do your policies and procedures ensure the flexibility of your organization by defining ways of adapting to changes in the sector and the environment?<br>▪ Have you established a good cooperation level with other sectoral organizations (inter-organizational strategic ties)? |
|  | Password Robustness and Exposure | ▪ Does your log-on procedure avoid displaying a password being entered?<br>▪ Are your computers set up so that others cannot view staff entering passwords? |
|  | Privileged Account Management | ▪ Do you identify the privileged access rights associated with each system or process and the users to whom they need to be allocated?<br>▪ Do you log changes to privileged accounts? |
|  | Role Segregation | ▪ Do you properly inform employees about his responsibilities that remain valid after termination or change of employment?<br>▪ Are access permissions and authorizations managed according to the principles of least privilege and separation of duties? |
|  | Third-Party Relationships | ▪ Have you formalized contractual relationships with partners and suppliers regarding information security?<br>▪ Do you identify and define the necessary requirements a third party should have to be considered trusty? |
|  | Wireless Access Management | ▪ Do you maintain an inventory of authorized wireless access points connected to the wired network?<br>▪ Have you created a separate wireless network for personal or untrusted devices? |
| **Operations** | Compliance Review | ▪ Do you audit your processes and procedures for compliance with established policies and standards?<br>▪ Do you review and revise your security documents, such as: policies, standards, procedures, and guidelines, on a regular basis? |

| | | | |
|---|---|---|---|
| | Documentation Fulfillness | ▪ | Do you have all the necessary policies and procedures properly documented? |
| | | ▪ | Do you have all the necessary records properly documented? |
| | Efficient Distinction of Development, Testing and Operational Environments | ▪ | Do users have different user profiles for operational and testing systems? |
| | | ▪ | Do you maintain separate environments for production and non-production systems? |
| | Operating Procedures | ▪ | Do you specify the operational instructions of the installation and configuration of systems? |
| | | ▪ | Do you specify the operational instructions of the scheduling requirements, including interdependencies with other systems, earliest job start and latest job completion times? |
| | Organizational Culture and Top Management Support | ▪ | Is your leadership actively and continuously involved in information security planning? |
| | | ▪ | Do you pursue the principle of efficiency in information security - economy/cost optimization? |
| | Risk Assessment | ▪ | Do you receive threat and vulnerability information from information sharing forums and sources? |
| | | ▪ | Is the organizational risk tolerance determined and clearly expressed? |
| **Defense** | Boundary Defense | ▪ | Do you maintain an up-to-date inventory of all of the organization's network boundaries? |
| | | ▪ | Do you decrypt all encrypted network traffic at the boundary proxy prior to analyze the content? |
| | Cryptography | ▪ | Do you encrypt all data stored in cloud services? |
| | | ▪ | Do you encrypt event files locally and in transit? |
| | Email and Web Browser Resilience | ▪ | What is the percentage from your total received emails that are detected as spam? |
| | | ▪ | What is the percentage of your SSL certificates that are configured incorrectly? |
| | Information Security Policy and Compliance | ▪ | Have you properly broken-down information security policies into sub-areas and orderly documented them? |
| | | ▪ | Do your policies and procedures comply with relevant regional legislation? |
| | Malware Defense | ▪ | What percentage of your systems (workstations, laptops, servers) are covered by antivirus/antispyware software? |
| | | ▪ | Do you send all malware detection events to enterprise anti-malware administration tools and event log servers for analysis and alerting? |
| | Security Awareness and Training Program | ▪ | Do you perform a skills gap analysis to understand the skills and behaviors workforce members are not adhering to, using this information to build a baseline education roadmap? |
| | | ▪ | Do you deliver training to address the skills gap identified to positively impact workforce members' security behavior? |
| **Security Governance** | Audit Logs Management | ▪ | Have you ensured that local logging has been enabled on all systems and networking devices? |
| | | ▪ | Do you protect logs from unauthorized alterations or deletion? |

| | Incident Response and Management | ▪ What percentage of your security incidents cause service interruption or reduced availability?<br>▪ Do you have established processes to receive, analyze and respond to vulnerabilities disclosed to the organization from internal and external sources (e.g. internal testing, security bulletins, or security researchers)? |
|---|---|---|
| | Penetration Tests and Red Team Exercises | ▪ Have you tested that you gracefully handle denial of service attempts (from compromised meters)?<br>▪ Do you apply a qualified third-party security penetration testing to test all hardware and software components prior to live deployment? |
| | Reporting Mechanisms | ▪ Do you provide your employees with a channel in order to report violations of information security policies or procedures?<br>▪ How much time does the organization take in order to respond to a report? |
| | Security Management Maturity | ▪ Are critical security tasks handled based on team decision-making techniques?<br>▪ Do you organize vertical and horizontal security meetings on a regular basis? |

Controls used by our evaluation methodology aim to assess whether specific security fields have been taken into consideration and to what extend rather than measure the effectiveness and efficiency of the actual policies and procedures in place. In other words, evaluate the multidisciplinary approach towards information security and the depths in which is organizationally reaching rather than the completeness of security technology solutions acquired and utilizedby the enterprise under examination.

This approach is even more evident in the individual level where the beliefs, emotions, attitude, and behavior of the employees is examined under various prisms using a variety of psychological, behavioral, emotional and specialization assessments.

Table 2. Individual Level presenting indicative controls

| Dimension | Domain | Indicative Controls |
|---|---|---|
| **Attitude** | Employee Climate | ▪ I believe that cyber criminals are more advanced than the people who are supposed to be protecting us.<br>▪ I worry that if I report a cyber-attack to the Police it might damage the reputation of the company. |
| | Employee Profiling | ▪ Seniority<br>▪ Enterprise role |
| | Employee Satisfaction | ▪ I am pleased with my organization's approach towards information security.<br>▪ I am happy to conform with the security guidance offered by our security experts. |
| **Awareness** | Policies and Procedures Awareness | ▪ Are you aware of the organizations' communication flows?<br>▪ Are you aware of the organization's role in the supply chain? |
| | Roles and Responsibilities Awareness | ▪ Are you aware of all the devices and systems you are responsible for?<br>▪ Are you aware of all the external information systems they come in contact with? |
| **Behaviour** | Policies and Procedures | ▪ Do you make sure your mobile devices are not left exposed? |

| | Compliance | ▪ Do you efficiently protect mobile devices from physical hazards? |
|---|---|---|
| | Security Agent Persona | ▪ What would you do if you saw a colleague not wearing their security pass around the office?<br>▪ What would you do if you overheard a discussion, which you knew to be about some highly sensitive and confidential information, being held in a corridor where external visitors often pass through? |
| | Security Behaviour | ▪ How many of your security incidents stem from non-secure behavior?<br>▪ I get into the office wearing my security pass. |
| **Competency** | Employee Competency | Specific per organization and employee. |
| | Security Skills Evaluation | ▪ What is necessary for a person to turn a plain text message into an encrypted message?<br>▪ Which of the following events presents the greatest risk? |
| | Training Completion and Scoring | ▪ My achievement score at the last security training program I participated in was around.<br>▪ How many self-security assessments do you normally attempt per year? |

## 4. DESIGNING THE SURVEY

Our goal was to design a survey that would be short and targeted to get the security pulse of current business reality in the critical infrastructure domain. One of our major aims was to keep the questionnaire small and easily addressed in a timely manner by a common employee with no special security expertise or knowledge. This way, we could facilitate participation of a broader workforce group lessening effort and prerequisites while maximizing result variation and credibility. Towards that goal, we needed to formulate questions targeting specific security factors bridging various security domains while smartly extracting information depicting the existing working security routine and culture, their disruption by the COVID-19 crisis and their reaction to these special and rather demanding circumstances.

On the other hand, taking into consideration the reported cyber-crime incidents along with the fraud and attack techniques used by the criminals of the dark web during this period, we focused our evaluation on specific dimensions related to network infrastructure, asset management, business continuity, employee awareness, and attitude.

In the paragraphs to follow, we outline how starting from a detailed cyber-security culture framework with more than 500 controls, we have narrowed down our objectives to a questionnaire containing no more than 23 questions, depending on the provided answers. Table 3indexes the questions constituting the final version of our questionnaire including secondary clarification questions presented based on provided participant input whereas Table 4correlates each of the questions to specific cyber-security levels, dimensions, and domains of our model.

Table 3. Question indexing, including secondary clarification questions presented based on provided input (asterisk annotated).

| Q1 | Prior to the COVID-19 crisis, were you able to work from home? | Q9.2 | How were you informed how to use them? | Q12.6 | I am proud to work for my organisation. |
|---|---|---|---|---|---|
| Q2.1 | Did you receive any security guidelines from your employer regarding working from home? | Q10.1 | Has your company adopted a specific collaboration solution? | Q12.7 | I have access to the things I need to do my job well. |
| Q2.2* | Please describe the main (2-3) security guidelines provided. | Q10.2* | What abilities does it offer? | Q13 | What is your age? |
| Q3 | What kind of devices are you using to connect to your corporate working environment? | Q11.1 | Did you face any of the below cyber-security related threats during the COVID-19 crisis? | Q14 | What is the highest degree or level of school you have completed? |
| Q4 | Are these devices accessed by users other than yourself? | Q11.2* | Please name any other cyber-security threats you encountered during this period, not listed above. | Q15 | Please select the business domain of the organisationyou work for. |
| Q5 | These devices are personal or corporate assets? | Q12.1 | To what extent do you agree with the following statements: (during this specific period of the COVID-19 crisis)  I prefer working from home than going to the office. | Q16 | Which of the following best describes your work position? |
| Q6 | Are these devices managed by your organisation? | Q12.2 | I work more productively from home. | Q17 | Comments |
| Q7 | Which of the following apply for the devices you currently use for your working from home employment? (providing security measures alternatives, e.g. antivirus, password protections) | Q12.3 | I collaborate with my colleagues as effectively as when we are in office. | | |
| Q8 | How do you obtain access to your corporate working environment? | Q12.4 | I am satisfied by my employer's approach to the crisis. | | |
| Q9.1 | Were you asked to use applications or | Q12.5 | I have all the support I need to face any | | |

| | services that you were unfamiliar with, because of the need for remote working? | technical problems I have (e.g. corporate access issues, infrastructure failures, etc.). | | |
|---|---|---|---|---|

## 4.1. Organizational Level

Culture is defined as a set of shared attitudes, values, goals, and practices that define an institution or organization. Consequently, cyber-security culture refers to the set of values, conventions, practices, knowledge, beliefs and behaviors associated with information security. Therefore, its skeleton is being outlined by the working environment along with the technological infrastructure and security countermeasures that define it.

To understand, assess and detail the security cultural status of the critical infrastructure organizations represented in our survey, we have included questions Q1-Q10 that heartbeat the overall technological and security readiness and adaptability. Under the coronavirus prism, we intended to understand if teleworking was an option prior to the crisis or not and under which security policies. Thus, we have included queries polling the remote access procedures and their meeting standards as well as the types, configuration and management of the devices used to gain access to the corporate environments. In other words, we attempted to assess the way and the means of the working from home experience with a clear focus on cyber-security.

Given that one of the most important shifts in the "everyday business" was the "home office", we clearly focused on the technological infrastructure used during this period. Preconfigured workstations were substituted by remote assets, such as laptops, tablets, smartphones, which gained access to the corporate network via different communication protocols and security standards. We needed to evaluate this newly deployed working ecosystem and understand the security precautions made by the participating enterprises.

Additionally, we intended to assess the security maturity of the management, the security response team, and awareness training program by introducing several questions clearly related to cyber-security familiarity and readiness. The most critical question of these category is the one referring to security guidelines provided during the COVID-19 crisis seeking to match their responsiveness and contemporary. Most of the leading cyber-security entities and experts during the coronavirus pandemic have issued special security guidelines towards individuals and organizations. Succeeding or failing to propagate similar guiding principles towards your workforce during such confusing and challenging periods is a core security indicator.

Another business facet which was examined, although not directly related to information security, was the collaboration possibilities offered to employees. Business trend and current reality (including but not limited to COVID-19 crisis) lead to a "remote" or "hybrid" working structure [23, 24]. Communication and teamwork need to be facilitated and promoted, especially during this period when general isolation is mandated as the main defense against the virus spread. Companies are expected to make provision for all means necessary to assist their employees in being productive, effective and cooperative [25]. This notion and quality are being tested via two simplified questions included in our survey.

## 4.2. Individual Level

Moving down to an individual level, evaluation becomes more demanding since virus fear and emotional stress dominate every aspect of daily life directly or indirectly affecting the human-related security factors. Questions Q11-Q12 attempt to probe the security behavior, attitude and competency of the remote workers by examining their emotions, thoughts and beliefs and by asking them to report any security incidents they came up against.

As Cisco CEO John Chambers stated in a January 2015 post for the World Economic Forum titled "What does the Internet of Everything mean for security?", "there are two types of companies: those who have been hacked, and those who don't yet know they have been hacked". Similarly, there are two types of technology users, those who understand the perils they face every day, either successfully or not, and those who are simply unaware of the dangers they are most probably already exposed to. Our question Q11 tries to examine this security familiarity and awareness among the different seniority, age, expertise, background of the participants.

Question Q12 clearly traces the emotional and intellectual state of the remote workers by listing several questions deriving from the Insider Threat theory according to which dissatisfaction, stressful events and personality predispositions can transform a loyal employee to a malicious or unintentional insider [26, 27].

Questions Q13-Q16 refer to generic information used for an individual profiling and categorization which shall enable us to analyze gathered results under different prisms offering various grouping possibilities and leading to possibly interesting findings on the basis of age, industry, education, working experience and expertise. This enterprise profiling is also believed to be closely related to security factors formulating fertile ground to a number of human-related cyber-threats [28, 29, 30].

Table 4. Correlating questions to cyber-security culture framework

| | Organizational Level | | | | | | Individual Level | | | |
|---|---|---|---|---|---|---|---|---|---|---|
| | Assets | Continuity | Access and Trust | Operations | Defense | Security Governance | Attitude | Awareness | Behavior | Competency |
| Q1 | - Network Infrastructure Management - Network Configuration Management | | - Access Management - External Environment Connections | | | | | | | |
| Q2.1 Q2.2* | | Change Management | | Organizational Culture and Top Management Support | Security Awareness and Training Program | Security Management Maturity | | | | |
| Q3 | Hardware Assets Management | | Access Management | | | | | | | |

| | | | | | | | | |
|---|---|---|---|---|---|---|---|---|
| | ment | | | | | | | |
| **Q4** | | | Access Management | | | | | - Policies and Procedures Compliance<br>- Security Behavior | |
| **Q5** | - Hardware Assets Management<br>- Information Resources Management<br>- Data Security and Privacy | | - Access Management<br>- External Environment Connections | | | | | | |
| **Q6** | - Hardware Assets Management<br>- Software Assets Management<br>- Information Resources Management<br>-Data Security and Privacy | | -Access Management<br>-External Environment Connections | | | | | | |
| **Q7** | -Hardware Configuration Management<br>-Information Resources Management<br>-Data Security and Privacy | | | | Malware Defense | | Policies and Procedures Awareness | - Policies and Procedures Compliance<br>- Security Behavior<br>- Security Agent Persona | |
| **Q8** | - Network Infrastructure Management<br>- Network Configuration Management | | - Access Management<br>- External Environment Connections | | Boundary Defense | | | | |
| **Q9.1** | | - Business Continuity & Disaster Recovery<br>- Change Management | | | | | | | |
| **Q9.2** | | | Communication | Organizational Culture and Top Management Support | | | | | |
| **Q10.1** | | | | Operating Procedures | | | | | |
| **Q10.2*** | | | | | | | | | |
| **Q11.1** | | | | | | | | -Security Behavior<br>-Security Agent Persona | Security Skills Evaluation |
| **Q11.2*** | | | | | | | | | |
| **Q12.1** | | | | | | Employee Climate | | | |
| **Q12.2** | | | | | | | | | |
| **Q12.3** | | | | | | | | | |
| **Q12.4** | | | | | | Employee Satisfaction | | | |
| **Q12.5** | | | | | | | | | |
| **Q12.6** | | | | | | | | | |
| **Q12.7** | | | | | | | | | |
| **Q13** | | | | | | Employee Profiling | | | |
| **Q14** | | | | | | | | | |
| **Q15** | | | | | | | | | |
| **Q16** | | | | | | | | | |

The accruing questionnaire manages to combine the two security levels of our framework effectively and efficiently. Additionally, its contents have been tailored to rapidly yet effectually heartbeat the cyber-security reality during a disrupting chronological period, such as the COVID-19 pandemic. This agile instrument, although offering a quick and fruitful measurement method compared to similar concurrent surveys, it cannot be considered an in-depth information security assessment. Furthermore, it should not be used to label participating organizations but only to provide an overview of the AS-IS situation.

Having developed a first questionnaire version addressing the security elements of interest based on our security culture framework, we carefully designed the rest of the survey methodology including:

- **validity testing:** identify ambiguous questions or wording, unclear instructions, or other problems before widespread dissemination possibly conducted by a group of survey experts, experienced researchers and analysts, certified security and technology officers.
- **delivery method:** select the appropriate delivery method and possibly run an instrument validity testing to verify survey conduction methodology.
- **sample selection:** carefully chose representatives from energy, transport, water, banking, financial market, healthcare and digital infrastructure from different European countries (e.g. Cyprus, France, Ger-many, Greece, Italy, Romania, Spain) affected by the COVID-19 crisis.
- **survey duration:** defining a specific start and end period communicated to all invited parties.

Survey has been concluded and its results have been made available via a research data repository [31]. Their analysis and interesting findings are currently under publication.

## 5. NEXT STEPS

Having performed this indicative(meaning non-restricted within the business limits of an organization) pilot application of our cyber-security culture framework during a crisis period, we now focus on conducting similar cyber-security culture assessment surveys to specific organizations targeting different security domains and not only during the pandemic. We intend to proceed with extended applications of our framework and conduct a comparative analysis on the cyber-security differences among the various business domains possibly identifying variations in needs, threats, attitude, awareness, and overall culture.

## 6. CONCLUSIONS AND FUTURE WORK

Our survey focuses on evaluating the security readiness and responsiveness of corporations during the Great Shutdown and more specifically it shall be addressing critical infrastructure domain representatives from different countries affected by the coronavirus.

Security cultural approach demands flexibility and concurrency. Corporations of the public and private sector regardless of their specialization and expertise need to develop a digital workplace strategy that includes collaboration applications, security controls, network management, and digital technologies. They need to adjust their employment modes, business models, communication, and marketing channels.

In a radically evolving and transforming environment, security and risk teams need to become part of the crisis management group, remote working employees need to remain vigilant to cyber-threats and operations life-cycle needs to remain uninterrupted especially for the operators of essentials services. Our research aims to investigate if and to what extend is this approach embraced by the critical infrastructure organizations in different countries nowadays while revealing interesting findings related to cyber-security and inspiring further scientific research on this field.

Since unfortunately COVID-19 crisis still holds, we now consider repeating our survey properly adjusted and addressed to the same organizations involved in the first iteration, to understand if their security culture status has differentiated and evolved via this crash-testing experience or if the long-standing critical circumstances have negatively affected their attitude and behavior towards information technology and security. We are also examining a similar tailor-made survey focusing on other domains of interest possibly differentiating organizations of the public and private sector and different financial or workforce dimensions.

## ACKNOWLEDGMENT

This project has received funding from the European Union's Horizon 2020 research and innovation programme under grant agreement No 832907 [32].

## REFERENCES

[1]   T. P. Velavan and C. G. Meyer, "The COVID-19 epidemic," Tropical Medicine and International Health, vol. 25, no. 3, p. 278–280, 2020.

[2]   D. S. Hui, E. I. Azhar, T. A. Madani, F. Ntoumi, R. Kock, O. Dar, G. Ippolito, T. D. Mchugh, Z. A. Memish, C. Drosten, A. Zumla and E. Petersen, "The continuing 2019-nCoV epidemic threat of novel coronaviruses to global health — The latest 2019 novel coronavirus outbreak in Wuhan, China," International Journal of Infectious Diseases, vol. 91, pp. 264-266, 2020.

[3]   "WHO Director-General's opening remarks at the media briefing on COVID-19," World Health Organization (WHO), 2020.

[4]   United Nations Development Programme, "Socio-economic impact of COVID-19 | UNDP," 2020. [Online]. Available: https://www.undp.org/content/undp/en/home/coronavirus/socio-economic-impact-of-covid-19.html. [Accessed 07 07 07].

[5]   M. Wolf, "The world economy is now collapsing," Financial Times, 14 04 2020. [Online]. Available: https://www.ft.com/content/d5f05b5c-7db8-11ea-8fdb-7ec06edeef84. [Accessed 01 07 2020].

[6]   E. Talamàs, "The Great Shutdown: Challenges And Opportunities," Forbes, 14 05 2020. [Online]. Available: https://www.forbes.com/sites/iese/2020/05/14/the-great-shutdown-challenges-and-opportunities/#60eaf6e86f12. [Accessed 07 07 2020].

[7]   G. Gopinath, "The Great Lockdown: Worst Economic Downturn Since the Great Depression," IMFBlog, 14 04 2020. [Online]. Available: https://blogs.imf.org/2020/04/14/the-great-lockdown-worst-economic-downturn-since-the-great-depression/. [Accessed 07 07 2020].

[8]   "ILO Monitor:COVID-19 and the world of work. Second edition," International Labour Organization (ILO), 2020.

[9]   A. G. Blanco, "The impact of COVID-19 on the spread of cybercrime," BBVA, 27 04 2020. [Online]. Available: https://www.bbva.com/en/the-impact-of-covid-19-on-the-spread-of-cybercrime/. [Accessed 07 07 2020].

[10]  "COVID-19 cyberthreats," INTERPOL, 2020. [Online]. Available: https://www.interpol.int/en/Crimes/Cybercrime/COVID-19-cyberthreats. [Accessed 07 07 2020].

[11]  "Cybercriminals targeting critical healthcare institutions with ransomware," INTERPOL, 04 04 2020. [Online]. Available: https://www.interpol.int/en/News-and-Events/News/2020/Cybercriminals-targeting-critical-healthcare-institutions-with-ransomware. [Accessed 07 07 2020].

[12]  C. D. Robert Richardson, "CSI Computer Crime and Security Survey 2010/2011," Computer Security Institute (CSI), 2011.

[13]  "2019 HIMSS Cybersecurity survey," Healthcare Information and Management Systems Society, 2019.

[14] T. Menze, "The State Of Industrial Cybersecurity," ARC Advisory Group, 2019.

[15] B. Filkins and D. Wylie, "SANS 2019 State of OT/ICS Cybersecurity Survey," SANS™ Institute, 2019.

[16] "The future of cyber survey 2019," Deloitte, 2019.

[17] "SMESEC - SMEs' Cybersecurity Watch," SMESEC , [Online]. Available: https://docs.google.com/forms/d/e/1FAIpQLSdLcZ_6LKFANf6oX2XlOf_4Q55yDGR7ZZKCus7md 4F1IsqvEQ/viewform. [Accessed 17 04 2020].

[18] Information Security Community, "Cybersecurity Trends Survey," [Online]. Available: https://www.surveymonkey.com/r/Cybersecurity-DR. [Accessed 17 04 2020].

[19] "COVID-19, Info Stealer & the Map of Threats – Threat Analysis Report," Reason Labs, 09 03 2020. [Online]. Available: https://blog.reasonsecurity.com/2020/03/09/covid-19-info-stealer-the-map-of-threats-threat-analysis-report/. [Accessed 16 04 2020].

[20] "Home working: preparing your organisation and staff," National Cyber Security Centre, 17 03 2020. [Online]. Available: https://www.ncsc.gov.uk/guidance/home-working. [Accessed 17 04 2020].

[21] "Working From Home - COVID19 - ENISA," ENISA - European Union Agency for Cybersecurity, [Online]. Available: https://www.enisa.europa.eu/topics/WFH-COVID19?tab=details. [Accessed 17 04 2020].

[22] A. Georgiadou, S. Mouzakitis, K. Bounas and D. Askounis, "A Cyber-Security Culture Framework for Assessing Organization Readiness," Journal of Computer Information Systems, 2020.

[23] Tessian, "Securing the Future of Hybrid Working," Tessian, 2020.

[24] "The 2020 State of Remote Work," Buffer & AngelList, 2020.

[25] S. Shen, J. Sun, D. D. Wan, A. Gao, L. Mok and O. Chen, "Coronavirus (COVID-19) Outbreak: Short- and Long-Term Actions for CIOs," Gartner, 2020.

[26] D. Cappelli, A. Moore and R. Trzeciak, The CERT Guide to Insider Threats: How to Prevent, Detect, and Respond to Information Technology Crimes (Theft, Sabotage, Fraud), Boston: Addison-Wesley Professional, 2012.

[27] CERT Insider Threat Team, "Unintentional Insider Threats: A Foundational Study," Software Engineering Insitute, Pittsburgh, 2013.

[28] S. R. Band, D. Cappelli, L. F. Fischer, A. P. Moore, E. D. Shaw and R. F. Trzeciak, "Comparing Insider IT Sabotage and Espionage: A Model-Based Analysis," Software Engineering Institute, Pittsburgh, 2006.

[29] M. Bishop, "Position: "insider" is relative," in Proceedings of the 2005 Workshop on New Security Paradigms, Lake Arrowhead, California, 2005.

[30] M. Bishop and C. Gates, "Defining the insider threat.," in Proceedings of the 4th annual workshop on Cyber security and information intelligence research: developing strategies to meet the cyber security and information intelligence challenges ahead, Oak Ridge Tennessee USA, 2008.

[31] A. Georgiadou, S. Mouzakitis and D. Askounis, "Working from home during COVID-19 crisis – A Cyber-Security Culture Assessment Survey," Mendeley Data, Athens, 2020.

[32] "Energy Shield," Energy Shield, 2019. [Online]. Available: https://energy-shield.eu/. [Accessed 25 03 2020].

# AUTHORS

**Mrs. Anna Georgiadou** is a research associate in the Management and Decision Support Systems Laboratory in the School of Electrical and Computer Engineering at the National Technical University of Athens (NTUA). She has been working as a senior engineer on operation and system support services for major telecommunication providers, energy regulators, banking systems, and other critical national infrastructures. She has recently become a Ph.D. candidate in the cyber-security field inspired by her active participation in the HEDNO's (Hellenic Electricity Distribution Network Operator) information security management group. She is a certified database and network administrator, software developer, and data analyst.

**Dr. Spiros Mouzakitis** is a senior research analyst for the National Technical University of Athens (NTUA). He has 18 years of industry experience in the conception, analysis, and implementation of information technology systems. His research is focused on decision analysis in the field of decision support systems, risk management, Big Data Analytics, as well as optimization systems and algorithms, and enterprise interoperability. He has published in numerous journals including Computer Standards & Interfaces, International Journal of Electronic Commerce, Information Systems Management, and Lecture Notes in Computer Science, and has presented his research at international conferences.

**Dr. Dimitris Askounis** is a Professor in the School of Electrical and Computer Engineering at the National Technical University of Athens (NTUA). He has been involved in numerous IT research and innovation projects funded by the EU since 1988 in the thematic areas of eBusiness interoperability, eGovernance, data exploitation and management, decision support, knowledge and quality management, computer integrated manufacturing, enterprise resource planning, etc. He teaches digital innovation management, decision support, and management systems, and he is a member of scientific committees on innovation and entrepreneurship competitions and incubators offered by International University networks, Financial Institutions, etc. Dr. Askounis has published over 80 papers in scientific journals and international conference proceedings.